\pdfoutput=1
\documentclass[12pt,a4paper]{article}

\usepackage{ifthen} 
\newboolean{pdflatex}
\setboolean{pdflatex}{true} 

\newboolean{articletitles}
\setboolean{articletitles}{true} 

\newboolean{uprightparticles}
\setboolean{uprightparticles}{false} 

\def\paperauthors{LHCb collaboration} 
\def\paperasciititle{Test of lepton universality using B->D(*)+tau-nutau decays} 
\def\papertitle{Measurement of the branching fraction ratios $R(D^{+})$ and $R(D^{*+})$ using muonic $\tau$ decays} 
\def\paperkeywords{{High Energy Physics}, {LHCb}} 
\def\papercopyright{\the\year\ CERN for the benefit of the LHCb collaboration} 
\def\paperlicence{CC BY 4.0 licence}
\def\paperlicenceurl{https://creativecommons.org/licenses/by/4.0/}

\usepackage[top=1in, bottom=1.25in, left=1in, right=1in]{geometry}

\usepackage{microtype}
\usepackage{lineno}  
\usepackage{xspace} 
\usepackage{caption} 

\usepackage{graphicx}  
\usepackage{color}
\usepackage{colortbl}
\graphicspath{{./figs/}} 

\usepackage{amsmath} 
\usepackage{amssymb}
\usepackage{amsfonts}
\usepackage{upgreek} 

\newcommand*\patchAmsMathEnvironmentForLineno[1]{%
\expandafter\let\csname old#1\expandafter\endcsname\csname #1\endcsname
\expandafter\let\csname oldend#1\expandafter\endcsname\csname
end#1\endcsname
 \renewenvironment{#1}%
   {\linenomath\csname old#1\endcsname}%
   {\csname oldend#1\endcsname\endlinenomath}%
}
\newcommand*\patchBothAmsMathEnvironmentsForLineno[1]{%
  \patchAmsMathEnvironmentForLineno{#1}%
  \patchAmsMathEnvironmentForLineno{#1*}%
}
\AtBeginDocument{%
\patchBothAmsMathEnvironmentsForLineno{equation}%
\patchBothAmsMathEnvironmentsForLineno{align}%
\patchBothAmsMathEnvironmentsForLineno{flalign}%
\patchBothAmsMathEnvironmentsForLineno{alignat}%
\patchBothAmsMathEnvironmentsForLineno{gather}%
\patchBothAmsMathEnvironmentsForLineno{multline}%
\patchBothAmsMathEnvironmentsForLineno{eqnarray}%
}

\usepackage{hyperxmp}

\usepackage[pdftex,
            pdfauthor={\paperauthors},
            pdftitle={\paperasciititle},
            pdfkeywords={\paperkeywords},
            pdfcopyright={Copyright (C) \papercopyright},
            pdflicenseurl={\paperlicenceurl}]{hyperref}

\usepackage[colorinlistoftodos,textsize=scriptsize]{todonotes}

\usepackage[bottom,flushmargin,hang,multiple]{footmisc}

\usepackage[all]{hypcap} 

\usepackage{xspace} 
\usepackage{upgreek}

\def\lhcb   {\mbox{LHCb}\xspace}

\def\MagUp {\mbox{\em Mag\kern -0.05em Up}\xspace}

\ifthenelse{\boolean{uprightparticles}}%
{

 \def\Pmu         {\ensuremath{\upmu}\xspace}                 
 \def\Pnu         {\ensuremath{\upnu}\xspace}                 
                  
 \def\Ppi         {\ensuremath{\uppi}\xspace}

 \def\Ppsi        {\ensuremath{\uppsi}\xspace}

 \def\PDelta      {\ensuremath{\Delta}\xspace}                 
 \def\PXi         {\ensuremath{\Xi}\xspace}                 
 \def\PLambda     {\ensuremath{\Lambda}\xspace}                 
 \def\PSigma      {\ensuremath{\Sigma}\xspace}                 
 \def\POmega      {\ensuremath{\Omega}\xspace}                 
 \def\PUpsilon    {\ensuremath{\Upsilon}\xspace}
 \let\oldPi\Pi
 \def\PPi         {\ensuremath{\oldPi}\xspace}

 \def\PB      {\ensuremath{\mathrm{B}}\xspace}                 
                  
 \def\PD      {\ensuremath{\mathrm{D}}\xspace}

 \def\PJ      {\ensuremath{\mathrm{J}}\xspace}                 
 \def\PK      {\ensuremath{\mathrm{K}}\xspace}

 \def\Pb      {\ensuremath{\mathrm{b}}\xspace}                 
 \def\Pc      {\ensuremath{\mathrm{c}}\xspace}

 \def\Pi      {\ensuremath{\mathrm{i}}\xspace}

 \def\Ps      {\ensuremath{\mathrm{s}}\xspace}

 \def\thebaroffset{0.0em}
}
{

 \def\Pmu         {\ensuremath{\mu}\xspace}                 
 \def\Pnu         {\ensuremath{\nu}\xspace}                 
                  
 \def\Ppi         {\ensuremath{\pi}\xspace}

 \def\Ppsi        {\ensuremath{\psi}\xspace}                 
                  
 \mathchardef\PDelta="7101
 \mathchardef\PXi="7104
 \mathchardef\PLambda="7103
 \mathchardef\PSigma="7106
 \mathchardef\POmega="710A
 \mathchardef\PUpsilon="7107
 \mathchardef\PPi="7105
                  
 \def\PB      {\ensuremath{B}\xspace}                 
                  
 \def\PD      {\ensuremath{D}\xspace}

 \def\PJ      {\ensuremath{J}\xspace}                 
 \def\PK      {\ensuremath{K}\xspace}

 \def\Pb      {\ensuremath{b}\xspace}                 
 \def\Pc      {\ensuremath{c}\xspace}

 \def\Pi      {\ensuremath{i}\xspace}

 \def\Ps      {\ensuremath{s}\xspace}

 \def\thebaroffset{0.18em}
}
\newcommand{\offsetoverline}[2][\thebaroffset]{\kern #1\overline{\kern -#1 #2}}%

\makeatletter
\ifcase \@ptsize \relax
  \newcommand{\miniscule}{\@setfontsize\miniscule{4}{5}}
\or
  \newcommand{\miniscule}{\@setfontsize\miniscule{5}{6}}
\or
  \newcommand{\miniscule}{\@setfontsize\miniscule{5}{6}}
\fi
\makeatother

\DeclareRobustCommand{\optbar}[1]{\shortstack{{\miniscule (\rule[.5ex]{1.25em}{.18mm})}
  \\ [-.7ex] $#1$}}

\def\mun        {{\ensuremath{\Pmu^-}}\xspace} 

\def\mumu       {{\ensuremath{\Pmu^+\Pmu^-}}\xspace}

\def\neub       {{\ensuremath{\overline{\Pnu}}}\xspace}

\def\neumb      {{\ensuremath{\neub_\mu}}\xspace}

\def\squark    {{\ensuremath{\Ps}}\xspace}

\def\cquark    {{\ensuremath{\Pc}}\xspace}

\def\bquark    {{\ensuremath{\Pb}}\xspace}

\def\pion   {{\ensuremath{\Ppi}}\xspace}
\def\piz    {{\ensuremath{\pion^0}}\xspace}
\def\pip    {{\ensuremath{\pion^+}}\xspace}

\def\kaon    {{\ensuremath{\PK}}\xspace}

\def\KorKbar {\kern \thebaroffset\optbar{\kern -\thebaroffset \PK}{}\xspace}
\def\Kz      {{\ensuremath{\kaon^0}}\xspace}

\def\Kp      {{\ensuremath{\kaon^+}}\xspace}
\def\Km      {{\ensuremath{\kaon^-}}\xspace}

\def\D       {{\ensuremath{\PD}}\xspace}

\def\DorDbar {\kern \thebaroffset\optbar{\kern -\thebaroffset \PD}\xspace}
\def\Dz      {{\ensuremath{\D^0}}\xspace}

\def\Dp      {{\ensuremath{\D^+}}\xspace}
\def\Dm      {{\ensuremath{\D^-}}\xspace}

\def\DpDm    {\ensuremath{\Dp {\kern -0.16em \Dm}}\xspace}

\def\Dstarp  {{\ensuremath{\D^{*+}}}\xspace}

\def\Ds      {{\ensuremath{\D^+_\squark}}\xspace}

\def\Dsm     {{\ensuremath{\D^-_\squark}}\xspace}

\def\B       {{\ensuremath{\PB}}\xspace}
\def\Bbar    {{\ensuremath{\offsetoverline{\PB}}}\xspace}
\def\Bb      {{\ensuremath{\Bbar}}\xspace}
\def\BorBbar {\kern \thebaroffset\optbar{\kern -\thebaroffset \PB}\xspace}
\def\Bz      {{\ensuremath{\B^0}}\xspace}
\def\Bzb     {{\ensuremath{\Bbar{}^0}}\xspace}
\def\Bd      {{\ensuremath{\B^0}}\xspace}

\def\BdorBdbar {\kern \thebaroffset\optbar{\kern -\thebaroffset \Bd}\xspace}
\def\Bu      {{\ensuremath{\B^+}}\xspace}
\def\Bub     {{\ensuremath{\B^-}}\xspace}
\def\Bp      {{\ensuremath{\Bu}}\xspace}

\def\Bs      {{\ensuremath{\B^0_\squark}}\xspace}
\def\Bsb     {{\ensuremath{\Bbar{}^0_\squark}}\xspace}
\def\BsorBsbar {\kern \thebaroffset\optbar{\kern -\thebaroffset \Bs}\xspace}

\def\jpsi     {{\ensuremath{{\PJ\mskip -3mu/\mskip -2mu\Ppsi}}}\xspace}

\def\Y#1S{\ensuremath{\PUpsilon{(#1S)}}\xspace}

\def\Lz          {{\ensuremath{\PLambda}}\xspace}

\def\LorLbar     {\kern \thebaroffset\optbar{\kern -\thebaroffset \PLambda}\xspace}

\def\Lc          {{\ensuremath{\Lz^+_\cquark}}\xspace}

\def\Lb           {{\ensuremath{\Lz^0_\bquark}}\xspace}

\def\to                 {\ensuremath{\rightarrow}\xspace}

\def\qsq       {{\ensuremath{q^2}}\xspace}

\def\AT#1     {\ensuremath{A_{\mathrm{T}}^{#1}}\xspace}           

\def\C#1      {\ensuremath{\mathcal{C}_{#1}}\xspace}                       
\def\Cp#1     {\ensuremath{\mathcal{C}_{#1}^{'}}\xspace}                    
\def\Ceff#1   {\ensuremath{\mathcal{C}_{#1}^{\mathrm{(eff)}}}\xspace}        
\def\Cpeff#1  {\ensuremath{\mathcal{C}_{#1}^{'\mathrm{(eff)}}}\xspace}       
\def\Ope#1    {\ensuremath{\mathcal{O}_{#1}}\xspace}                       
\def\Opep#1   {\ensuremath{\mathcal{O}_{#1}^{'}}\xspace}                    

\newcommand{\aunit}[1]{\ensuremath{\text{\,#1}}}       

\newcommand{\tev}{\aunit{Te\kern -0.1em V}\xspace}
\newcommand{\gev}{\aunit{Ge\kern -0.1em V}\xspace}
\newcommand{\mev}{\aunit{Me\kern -0.1em V}\xspace}
\newcommand{\kev}{\aunit{ke\kern -0.1em V}\xspace}
\newcommand{\ev}{\aunit{e\kern -0.1em V}\xspace}
 
\newcommand{\mevc}{\ensuremath{\aunit{Me\kern -0.1em V\!/}c}\xspace}
\newcommand{\gevc}{\ensuremath{\aunit{Ge\kern -0.1em V\!/}c}\xspace}
\newcommand{\mevcc}{\ensuremath{\aunit{Me\kern -0.1em V\!/}c^2}\xspace}
\newcommand{\gevcc}{\ensuremath{\aunit{Ge\kern -0.1em V\!/}c^2}\xspace}
\newcommand{\gevgevcccc}{\ensuremath{\gev^2\!/c^4}\xspace} 

\def\fb   {\ensuremath{\aunit{fb}}\xspace}
\def\invfb   {\ensuremath{\fb^{-1}}\xspace}

\def\gsim{{~\raise.15em\hbox{$>$}\kern-.85em
          \lower.35em\hbox{$\sim$}~}\xspace}
\def\lsim{{~\raise.15em\hbox{$<$}\kern-.85em
          \lower.35em\hbox{$\sim$}~}\xspace}

\def\sPlot{\mbox{\em sPlot}\xspace}

\def\et         {\ensuremath{E_{\mathrm{T}}}\xspace}

\def\evtgen     {\mbox{\textsc{EvtGen}}\xspace}

\def\geant      {\mbox{\textsc{Geant4}}\xspace}

\def\photos     {\mbox{\textsc{Photos}}\xspace}

\def\pythia     {\mbox{\textsc{Pythia}}\xspace}

\def\tell1  {TELL1\xspace}
\def\ukl1   {UKL1\xspace}

\newcommand{\lhcborcid}[1]{\href{https://orcid.org/#1}{\hspace*{0.1em}\raisebox{-0.45ex}{\includegraphics[width=1em]{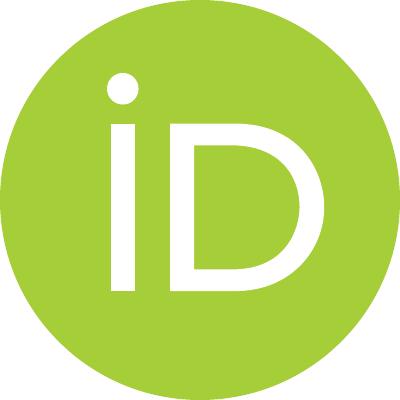}}}}

\usepackage{cite} 
\usepackage{mciteplus}

\def\DDstTauNu{\ensuremath{\Bbar\to D^{(*)} \tau^{-}\overline{\nu}_\tau\,}\xspace}
\def\DpTauNu{\ensuremath{\Bzb\to \Dp \tau^{-}\overline{\nu}_\tau\,}\xspace}
\def\DstpTauNu{\ensuremath{\Bzb\to \Dstarp \tau^{-}\overline{\nu}_\tau}\xspace}

\def\DDstMuNu{\ensuremath{\Bbar \to D^{(*)} \mu^{-}\overline{\nu}_\mu\,}}

\def\tautomu{\ensuremath{\tau^- \to \mun \neumb \nu_{\tau} \,}}
\def\DKpipi{\ensuremath{\Dp \to \Km\pip\pip \,}}
\def\El{\ensuremath{E_{\mu}^{*} \,}}
\def\msqmiss{\ensuremath{m_{\rm miss}^2 \,}}
\def\RDp{\ensuremath{R(D^{+})\,}}

\def\RDst{\ensuremath{R(D^{*})\,}}
\def\RDstp{\ensuremath{R(D^{*+})\,}}
\def\RDpDst{\ensuremath{R(D^{(*)+})\,}}
\def\RDpresult{\ensuremath{0.249 \pm 0.043 \pm 0.047}}
\def\RDstresult{\ensuremath{0.402 \pm 0.081\pm0.085}}
\def\corr{$-0.39$\;}
\def\compatibility{\ensuremath{0.78\,\sigma\,}}

\begin{document}

\renewcommand{\thefootnote}{\fnsymbol{footnote}}
\setcounter{footnote}{1}


\begin{titlepage}
  \pagenumbering{roman}
  
  \vspace*{-1.5cm}
  \centerline{\large EUROPEAN ORGANIZATION FOR NUCLEAR RESEARCH (CERN)}
  \vspace*{1.5cm}
  \noindent
  \begin{tabular*}{\linewidth}{lc@{\extracolsep{\fill}}r@{\extracolsep{0pt}}}
  \ifthenelse{\boolean{pdflatex}}
  {\vspace*{-1.5cm}\mbox{\!\!\!\includegraphics[width=.14\textwidth]{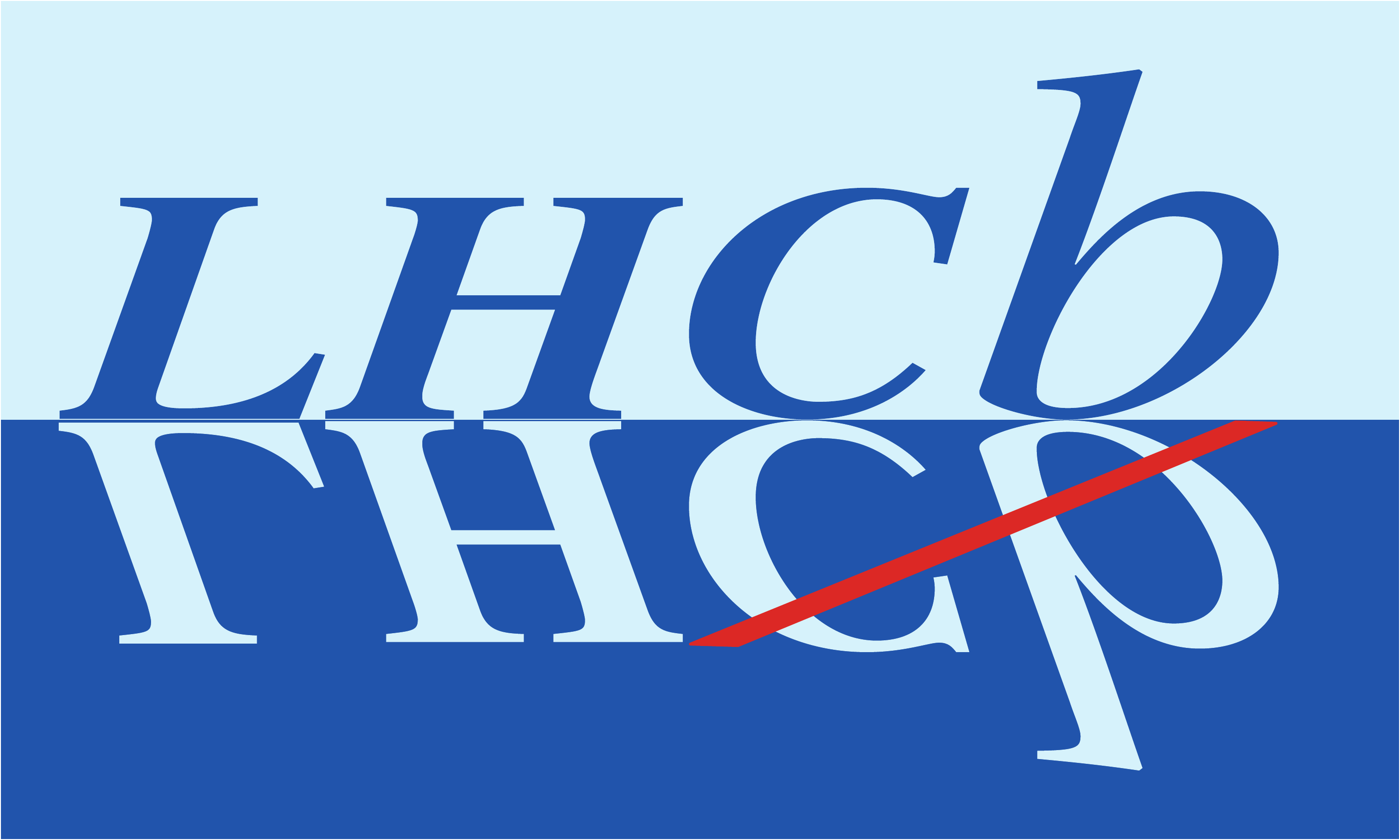}} & &}%
  {\vspace*{-1.2cm}\mbox{\!\!\!\includegraphics[width=.12\textwidth]{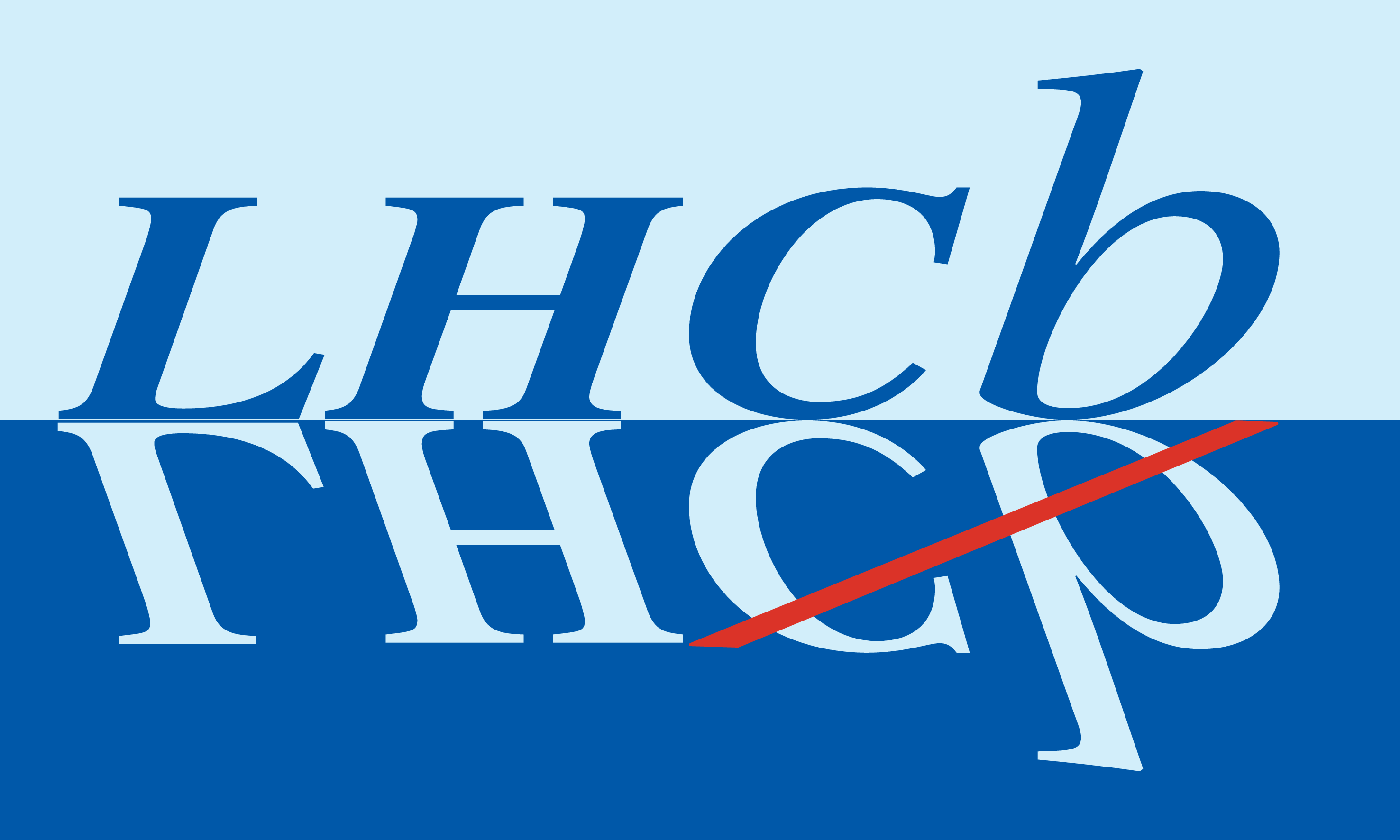}} & &}%
  \\
   & & CERN-EP-2024-125 \\  
   & & LHCb-PAPER-2024-007 \\  
   & & February 16, 2025 \\ 
   & & \\
  \end{tabular*}
  
  \vspace*{3.5cm}
  
  {\normalfont\bfseries\boldmath\huge
  \begin{center}
    \papertitle 
  \end{center}
  }
  
  \vspace*{1.5cm}
  
  \begin{center}
  \paperauthors\footnote{Authors are listed at the end of this letter.}
  \end{center}
  
  \vspace{\fill}
  
  \begin{abstract}
    \noindent
    The branching fraction ratios of $\kern 0.18em \overline{\kern -0.18em B}{}^0\to D^+\tau^-\overline{\nu}_{\tau}$ and $\kern 0.18em \overline{\kern -0.18em B}{}^0\to D^{*+}\tau^-\overline{\nu}_{\tau}$ decays are measured with respect to their muonic counterparts, using a data sample corresponding to an integrated luminosity of 2.0 fb$^{-1}$ collected by the LHCb experiment in proton-proton collisions at $\sqrt{s} = 13\,\text{Te\kern -0.1em V}$. The reconstructed final states are formed by combining $D^+$ mesons with $\tau^-\to\mu^-\overline{\nu}_{\mu}\nu_{\tau}$ candidates, where the $D^+$ is reconstructed via the $D^+\to K^-\pi^+\pi^+$ decay. The decay modes involving a \Dstarp meson are accessed via the decay $\Dstarp\to \Dp\piz$, where the \piz candidate is not reconstructed.  The results are
    \begin{align*}
    R(D^{+}) &= 0.249 \pm 0.043 \pm 0.047, \\
    R(D^{*+}) &= 0.402 \pm 0.081\pm 0.085,
    \end{align*}
    where the first uncertainties are statistical and the second systematic. The two measurements have a correlation coefficient of $-0.39$ and are compatible with the Standard Model.
  \end{abstract}
  
  \vspace*{1.5cm}
  
  \begin{center}
    Published in Phys.~Rev.~Lett. 134 (2025) 061801
  \end{center}
  
  \vspace{\fill}
  
  {\footnotesize 
  \centerline{\copyright~\papercopyright. \href{\paperlicenceurl}{\paperlicence}.}}
  \vspace*{2mm}
  
  \end{titlepage}

  
  \newpage
  \setcounter{page}{2}
  \mbox{~}

\renewcommand{\thefootnote}{\arabic{footnote}}
\setcounter{footnote}{0}

\cleardoublepage

\pagestyle{plain} 
\setcounter{page}{1}
\pagenumbering{arabic}

In the Standard Model (SM), the transition of a beauty quark ($b$) into a charm quark ($c$), a tau lepton ($\tau^-$) and a tau anti-neutrino ($\overline{\nu}_\tau$), $b \to c \tau^{-} \overline{\nu}_\tau$, proceeds via a tree-level $W$ boson exchange.\footnote{The inclusion of charge conjugate processes in this Letter is implied throughout.} The SM decay amplitude is large, resulting in large branching fractions and copious signal yields. Despite this, $b \to c \tau^{-} \overline{\nu}_\tau$ transitions are highly sensitive to new physics as the process involves three third-generation fermions~\cite{Fajfer:2012vx,Crivellin:2012ye,Tanaka:2012nw,Freytsis:2015qca}. This system therefore represents one of the best potential sources of contributions from new physics mediators that couple preferentially to the third generation~\cite{Allwicher:2023shc,Davighi:2023iks}. Furthermore, being a charged-current semileptonic transition, the decay amplitude can be factorised between the leptonic and hadronic parts to a good approximation. Uncertainties due to the strong interaction can be reduced by forming lepton universality ratios, where a $b \to c \tau^{-} \overline{\nu}_\tau$ transition is normalised to the corresponding muonic one, $b \to c \mu^{-} \overline{\nu}_\mu$. In the SM, these ratios only deviate from unity due to the different lepton masses.

The experimental challenge in analysing this system is the presence of multiple neutrinos in the final state, mandating a partial reconstruction of both the $\tau$-lepton and $b$-hadron decays. This partial reconstruction leaves the signal highly susceptible to a large number of backgrounds, which must be carefully controlled.
Previous analyses of the \mbox{$b \to c \tau^{-} \overline{\nu}_\tau$} system include measurements of the $\tau$-lepton and charm-hadron polarisation~\cite{Belle:2017ilt,LHCb:2023ssl} and the lepton universality ratios $R(D)$ and \RDst, which are formed from the branching fraction ratios ${\cal B}(\DDstTauNu)/{\cal B}(\DDstMuNu)$. Lepton universality is the SM symmetry of equal coupling to the different charged lepton species, only broken by the different masses of the leptons. A combination~\cite{HFLAV21} of the measurements~\cite{Lees:2012xj,Lees:2013uzd,Sato:2016svk,Huschle:2015rga,Belle:2019rba,Hirose:2017dxl,LHCb-PAPER-2022-052,LHCb-PAPER-2022-039} differs by approximately three standard deviations from the SM predictions~\cite{Bigi:2016mdz,Gambino:2019sif,Bordone:2019vic,Bernlochner:2017jka,Jaiswal:2017rve,BaBar:2019vpl,Martinelli:2021onb,Bigi:2017jbd,FermilabLattice:2021cdg}, indicating a potential enhancement of the semitauonic transition rate. Additional measurements of this process are therefore highly motivated.

This Letter describes the first simultaneous measurement of the lepton universality ratios $\RDpDst$ at the LHCb experiment, using the decays \tautomu, $\Dstarp\to \Dp\piz$ and $\DKpipi$, specifically the first LHCb measurement of $R(D^+)$. This measurement uses data corresponding to 2.0\invfb collected during the period 2015\textendash2016 in 13\tev $pp$ collisions. Using isospin symmetry, these measurements can be combined with those of $R(D^{(*)0})$. The analysis strategy is similar to the existing measurement performed with the $D^{0}\to K^{-}\pi^{+}$ decay~\cite{LHCb-PAPER-2022-039}, where a three-dimensional template fit to the squared four-momentum transferred to the lepton pair (\qsq), the energy of the muon in the $B$ meson rest frame (\El) and the squared invariant mass missing from the visible system (\msqmiss) is performed to separate the signal and the normalisation from the backgrounds. The contribution from the first excited states $D^{*}$ ($D^*(2007)^0$ and $D^*(2010)^+$) to the ground state $\Dp$ is significantly reduced in this measurement compared to the \Dz case in Ref.~\cite{LHCb-PAPER-2022-039}. The \piz meson in the $\Dstarp\to \Dp \piz$ decay is not reconstructed, due to the large background from neutral objects.

The \lhcb detector, described in detail in Refs.~\cite{LHCb-DP-2008-001,LHCb-DP-2014-002}, is a single-arm forward
spectrometer covering the \mbox{pseudorapidity} range $2<\eta <5$,
designed for the study of particles containing \bquark or \cquark
quarks. 
The online event selection is performed by a trigger~\cite{LHCb-DP-2012-004}, 
which consists of a hardware stage, based on information from the calorimeter and muon
systems, followed by a software stage, which applies a full event
reconstruction. 

Simulated events are used to optimise the signal selection, model the signal and backgrounds, and calculate the relative efficiency between the signal and normalisation channels. In the simulation, $pp$ collisions are generated using \pythia~\cite{Sjostrand:2006za,*Sjostrand:2007gs} with a specific \lhcb configuration~\cite{LHCb-PROC-2010-056}. Decays of hadronic particles are described by \evtgen~\cite{Lange:2001uf}, in which final-state radiation is generated using \photos~\cite{Golonka:2005pn}. The interaction of the generated particles with the detector, and its response, are implemented using the \geant toolkit~\cite{Allison:2006ve, *Agostinelli:2002hh} as described in Ref.~\cite{LHCb-PROC-2011-006}. For this Letter, a new fast simulation technique is used whereby the photon propagation in the Ring Imaging CHerenkov (RICH) detectors, shower development in the calorimeters and activity in the muon stations are not simulated. This tracker-only simulation is a factor of eight faster and has a 40\% smaller event size than full simulation. 
A smaller sample of simulated events which includes the full set of interactions (full simulation) is used for a cross-check and to develop specific tools such as those to reject partially reconstructed backgrounds with an additional neutral pion or photon in the final state.


At the hardware stage, events are required to be triggered by activity unrelated to the signal candidate or by high transverse energy signals in the HCAL due to the \Dp decay products. In the software stage, one- or two-track combinations of the $D^{+}$ decay products are required to pass a multivariate selection based on the transverse momentum and impact parameter with respect to an associated primary vertex (PV). A subsequent software level selects a displaced three-track combination with a $\Km\pip\pip$ invariant mass compatible with the \Dp mass and large transverse momentum. The $D^{+}$ candidate is combined with a muon with a loose vertex-quality requirement. The muon is required to have a momentum above 3\gevc to ensure that it reaches the muon stations. Particle identification (PID) requirements are applied on the final state particles, based on information from the RICH and muon systems.

A set of offline selection critera, following the final trigger stage selection, is additionally applied. The $D^{+}$ candidate is required to have a significant impact parameter with respect to any PV, which reduces background from charm decays originating directly from the $pp$ collision. A neural network classifier~\cite{DeCian:2255039} is also applied to all final-state particles to reject track candidates which have more than 30\% of hits that do not originate from the same source (fake tracks).

Background for which the \Dp candidate is an accidental track combination from different sources is rejected by employing a Boosted Decision Tree (BDT)~\cite{Breiman,AdaBoost} implemented in the TMVA
toolkit~\cite{Hocker:2007ht,*TMVA4}. The BDT is trained to distinguish simulated \DpTauNu decays as a proxy for the semitauonic signal from background, which is modelled using data selected with $\Km\pip\pip$ invariant mass outside the range 1840--1900\mevcc. The BDT features include kinematic and geometric information about the \Dp candidate as well as track quality variables for the decay products. The BDT requirement is optimised by maximising the metric $S/\sqrt{S+B}$, where $S$ is the expected number of signal decays from simulation and $B$ is the amount of background extrapolated from the mass sidebands in the data. This requirement is 97\% efficient for the signal and rejects 35\% of the background. 

Although the decay \DstpTauNu, where $D^{*+}\to D^{+}\piz$, is a signal channel in this analysis, the selection is optimised for the \DpTauNu decay and therefore an isolation requirement is imposed based on neutral calorimeter objects, typically produced in partially-reconstructed background decays. A dedicated BDT is trained to discriminate between \DpTauNu and \DstpTauNu decays, which is referred to as neutral isolation. This BDT uses the presence of neutral calorimeter objects in a cone around the \Dp direction and reconstructed $\piz\to\gamma\gamma$ candidates that have an invariant mass consistent with the known $\Dstarp$ mass~\cite{PDG2022}. A requirement on the output of this BDT is used to remove 30\% of \DstpTauNu candidates at the expense of 10\% \DpTauNu. The marginal gain is due to the large amount of activity in the ECAL from background processes.

Finally, a BDT-based isolation tool for charged tracks~\cite{LHCb-PAPER-2015-025}, referred to as track isolation, is used to classify all additional tracks in the event based on whether they are consistent with originating from the signal $B$-decay vertex. The three tracks that are most likely to originate from the signal candidate are used to control partially reconstructed backgrounds via a set of requirements on the BDT response that define isolation regions, described below.
 
A maximum-likelihood fit to the $\Km\pip\pip$ mass distribution of the selected candidates is performed, where the $\Dp$ signal is parameterised by an Hypatia function~\cite{MartinezSantos:2013ltf} and the background by an exponential function.
The resulting yields of $\Bbar\to D^{+}\mu^{-}\neumb X$ and background are  3.1M  and 150K, respectively. The fit also determines per-event signal weights via the \sPlot technique~\cite{Pivk:2004ty} that are used in the subsequent fit to the kinematic fit variables, \El, \qsq and \msqmiss, to subtract background where the $\Km\pip\pip$ combination does not originate from a \Dp meson.

Residual contributions of backgrounds with misidentified hadrons, such as \mbox{$\Ds\to\Kp\Km\pip$}, with a kaon misidentified as a pion, or \mbox{$\Lc\to p \Km\pip$} decays, with a proton misidentified as a pion, are cross-checked by computing the charm-hadron mass under the relevant mass hypotheses and validating that no visible signals are seen.


Several corrections are applied to improve the accuracy of the simulation. Firstly, the trigger and neutral isolation selection must be emulated for the tracker-only simulation. The hardware-level trigger must be emulated as it relies on information missing in the track-only simulation and is done by reproducing the two ways in which the events can be triggered. The efficiency for triggering via other particles in the event, which accounts for around 90\% of the candidates, is measured using $\Bp\to(\jpsi\to\mumu)\Kp$ decays as a function of the $\Bp$ kinematics. This is done directly using data in order to improve the description of the simulation.
The emulation of the efficiency for one of the \Dp decay products to directly trigger through a high \et cluster in the HCAL is determined with fully simulated signal decays as a function of the kinematics of the $\Dp\to\Km\pip\pip$ decay products.

The software trigger is accurately emulated by applying the same requirements offline, with the exception of the PID response, which is corrected for separately. The efficiency of the neutral isolation BDT requirement is modelled by applying a per-event weight, computed using a multivariate reweighter~\cite{BDTreweighter} trained on full simulation. With this approach, the efficiency can be parameterised as a function of the generator-level kinematic properties of the \Dp candidate, the sum of the four-momenta of the photons in the event, and the fit variables. The efficiencies of the PID requirements on the reconstructed final-state particles are determined using control samples~\cite{LHCb-DP-2018-001} as a function of track kinematics and the number of tracks in each event. They are also used to determine misidentification probabilities, which are crucial to estimate the background where the muon candidate is a misidentified track. The tracking efficiency is determined using $\jpsi\to\mumu$ decays in data~\cite{LHCb-DP-2013-002}, as a function of the track kinematics. The kinematics of the \B meson as well as the event multiplicity are corrected using $\Bp\to\jpsi\Kp$ decays.

Finally, a weight-based calibration procedure is designed to correct a comprehensive set of geometric and kinematic variables associated with the final-state particles as well as the Dalitz variables of the $\Dp\to\Km\pip\pip$ decay. In addition to the properties of the candidate, the kinematics of the three additional tracks found by the track-isolation algorithm are also corrected. This procedure is designed to correct for residual mismodelling of the simulation, such as the trigger response. The calibration sample for this correction procedure is selected candidates in the region $\msqmiss< 2\gevgevcccc$, where the sample is enriched in the normalisation channels. The result of this procedure depends on the relative abundance of the different decays assumed in the simulation, which is estimated by an initial fit to the data sample. The corrected simulation is consistent with the data in the distributions of all considered variables.

The corrected simulation is used to determine the relative efficiency between the signal and normalisation channels. This efficiency ratio is found to be 0.57 for the \DpTauNu mode and 0.58 for the \DstpTauNu mode. These ratios are smaller than unity since the relatively low momentum muon from the $\tau$ lepton has a small detection efficiency, and the $\tau$ lifetime results in a poorly defined $\Bz$ vertex. Systematic uncertainties on this ratio are subdominant compared to those affecting the signal yield determination.


The tauonic signal decay is only 2\% of the selected sample, split evenly between the \DpTauNu and \DstpTauNu modes, meaning that control over the background is essential to obtain a precise and accurate measurement. A three-dimensional binned-likelihood fit to \qsq, \El and \msqmiss is therefore employed to fully exploit the kinematic differences in the signal and background channels. Each contribution to the fit is modelled using a template histogram derived either from simulation or from specifically selected control samples in data. 

The data sample is split into four regions based on the response of the track isolation tool.  The first, referred to as the signal region, is selected by requiring that no track is in the vicinity of the signal candidate. Two additional samples with either exactly one or two additional charged pions, referred to as the one-pion and two-pion regions, are used to constrain the model parameters of the background involving $D^{**}$ states, where $D^{**}$ represents an excited charm meson that is heavier than the first excited states ($D^*(2007)^+$ and $D^*(2010)^0$). Finally, a sample with an additional charged kaon, referred to as the one-kaon region, is selected to control background from $\Bbar\to \Dp (X_{c} \to \mun X) (X')$ decays, where $X_c$ is a charm hadron, which often produces charged kaons, and $X$ represents any number of particles that are not reconstructed. These isolation regions are fitted simultaneously in order to propagate the corrections obtained from the control regions to the signal region. 

Uncertainties in the modelling of the components are incorporated into the fit as bin-by-bin correlated variations in the templates. These variations are implemented via a polynomial interpolation~\cite{Cranmer:1456844}. Exceptions to this are variations in the form factors, which are performed using the HAMMER tool~\cite{Bernlochner:2020tfi} as implemented in Ref.~\cite{GarciaPardinas:2020yrd}. The HAMMER tool allows for a fast, but exact, variation of a template as a function of the decay-model parameters. This variation is important to allow the form factors of both the signal and normalisation channels to vary as the constraints derived from precise lattice calculations~\cite{Na:2015kha,FermilabLattice:2021cdg,Harrison:2023dzh,PhysRevD.109.074503} can have larger uncertainties than those obtained from the fit. The form factors for the $\Bb\to\Dp$ and $\Bb\to\Dstarp$ modes are parameterised according to the BGL formalism~\cite{Boyd:1997kz} with multivariate Gaussian constraints applied according to the predictions reported in Refs.~\cite{PhysRevD.94.094008,FermilabLattice:2021cdg}.

The background originating from $\Bbar\to D^{**}\mun\neumb$ decays is less well-understood compared to those involving the lower mass states, despite a recent measurement by Belle~\cite{Belle:2022yzd}. This is because the measurements are less precise and the narrow-width assumption is less well-motivated due to the significant natural width of the $D^{**}$ states. The form-factor models employed for the $\Bbar \to D^{**}\mun\neumb$ decays are those described in Ref.~\cite{Bernlochner:2016bci}, with the parameters allowed to vary with Gaussian constraints applied according to the corresponding predictions. The tauonic components $\Bbar \to D^{**}\tau^{-}\bar{\nu}_{\tau}$ are constrained to the branching fraction predictions in Ref.~\cite{Bernlochner:2016bci} with a relative width of 50\% to the prediction values. Similar to Ref.~\cite{LHCb-PAPER-2022-039}, the modelling of this background is tuned and validated by fitting the one- and two-pion control regions. The former is used to determine the relative contribution of the lightest $D^{**}$ states ($D_{1}(2420)$, $D^{*}_{2}(2460)$, $D'_{1}(2430)$, and $D^{*}_{0}(2300)$) and the latter is used to determine the shape of heavier $D^{**}$ states such as the $D_{3}(2740)$ and $D(3000)^{0}$ mesons. As decays into the heavier states are particularly poorly known, a shape variation that depends linearly on the true \qsq of the decay is allowed to freely vary in the fit. The associated nuisance parameters are allowed to be different between the isolation regions in order to absorb the contribution of $\Lb\to \Dp n \mun\neumb$ decays. 

The most signal-like background is that from $\Bbar\to \Dp (X_{c}\to \mun X) (X')$ decays, where $X$ and $X'$ represent any number of particles originating from the $X_{c}$ or \Bbar hadrons, since a muon originating from a charm hadron has similar kinematics to that originating from a $\tau$ decay. This background is modelled using simulation with shape corrections allowed to vary in the fit, for which the one-kaon control region is crucial. The fit model includes two distinct categories of this background. The first arises from two-body transitions, such as the decay $\Bzb \to \Dp \Dsm$, and the second arises from multi-body transitions, such as the decay $\Bzb \to \Dp \Dm \Kz$. The two-body background is well reproduced by simulation whereas the multi-body modes are less well-understood, and a shape correction is therefore applied. The main correction consists of a pair of weights computed as linear and quadratic functions of the mass of the two charm hadrons, i.e. an empirical model similar to Ref.~\cite{LHCb-PAPER-2022-039}. Corrections to the relative size of the \Bub and \Bzb contributions to this background as well as the fraction of two-body to multi-body backgrounds are shared between all isolation regions. After the correction, a good fit quality in the one-kaon region is obtained, validating the procedure.

Unlike Ref.~\cite{LHCb-PAPER-2022-039}, the shape correction obtained in the one-kaon region is not applied directly to the signal region as a significant difference is observed in the parameters obtained from the two regions. This is not surprising as the Dalitz distribution of $\Bbar\to X_{c} \overline{X}_{c} K^{(*)}$ decays depends on the total charge of the $X_{c}\overline{X}_{c}$ combination, and the isolation requirements preferentially select certain charge combinations due to correlations with the presence of a \Kp meson. A separate correction based on the same parameterisation is applied for the signal region to allow for differences in the two regions. 

Background can also arise when a hadron track is misidentified as the muon candidate. This background arises mainly from $\Bbar \to \Dp h^{-} X$ decays, which have a total branching fraction of around 23\%, and is composed of hundreds of different individual decay modes, many of which have not been measured. Therefore, an accurate simulation of this background is unfeasible and a data-driven approach similar to Ref.~\cite{LHCb-PAPER-2022-039} is used. In this approach, a data sample is selected with identical requirements to the signal apart from a reversal of the muon-identification criteria. Data-driven misidentication efficiencies obtained from control samples are then used to translate the shape and size of this background to that expected in the signal region. 

A complication is that the misidentification sample consists of several different particle types, each with a different probability to be misidentified as a muon. In order to estimate the relative mixture of particle types, the misidentification sample is split into regions based on the PID response. The criteria for each region is designed such that the region is dominated by a particular particle type. For example, an electron misidentification region is determined by requiring that the particle track is identified as an electron based on the calorimeter and RICH systems. Cross-feed between the regions is determined by fitting the distribution of the number of candidates in each region. Templates for each particle type are determined from dedicated control samples in data with the exception of fake tracks, which are obtained from simulation. Once the relative abundance of each particle type is determined as a function of the track kinematics, the control samples are used to extrapolate the background into the signal region. Furthermore, a smearing of the track momentum~\cite{LHCb-PAPER-2022-039} is applied to account for the decay in flight of pions and kaons into muons, which has a small but statistically significant impact on the template shape.

Combinatorial background, where the $D^{+}$ and muon candidate do not originate from the same decay chain, is estimated from a sample where the \Dp is combined with a muon candidate of the same charge. This same-sign sample has no backgrounds originating from a single decay, apart from those where the muon is a misidentified track which are accounted for using the misidentification procedure described above. Possible differences in the same-sign sample and the combinatorial background are investigated by selecting a data sample where the $\Dp\mu^{\pm}$ invariant mass is greater than the mass of the \Bzb meson. This selects a pure sample of random combinations that can be compared to a same-sign sample selected in the same region. Multiple kinematic variables in this high mass region are then compared between the combinatorial and the same-sign samples. This comparison is used to derive a correction to the same-sign sample that is selected in the low visible-mass region.

\begin{figure}[tb]
    \centering
    \includegraphics[width=0.45\textwidth]{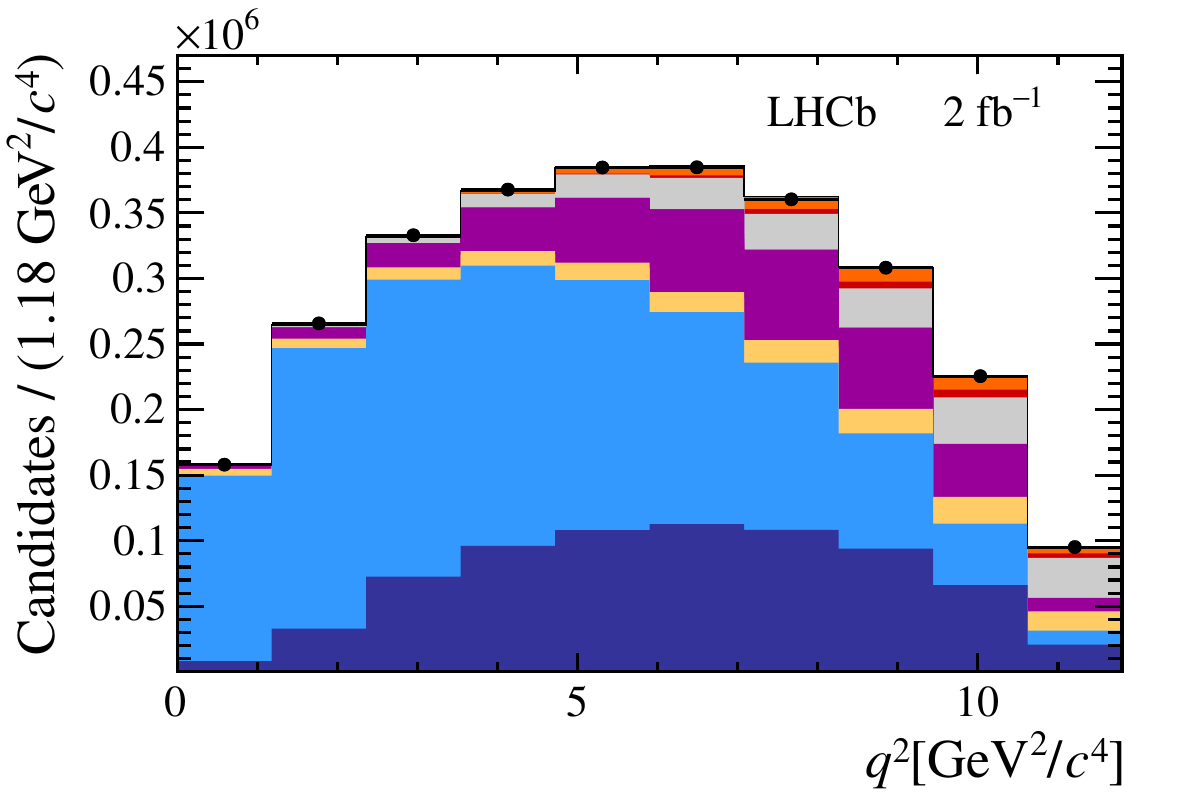}
    \includegraphics[width=0.45\textwidth]{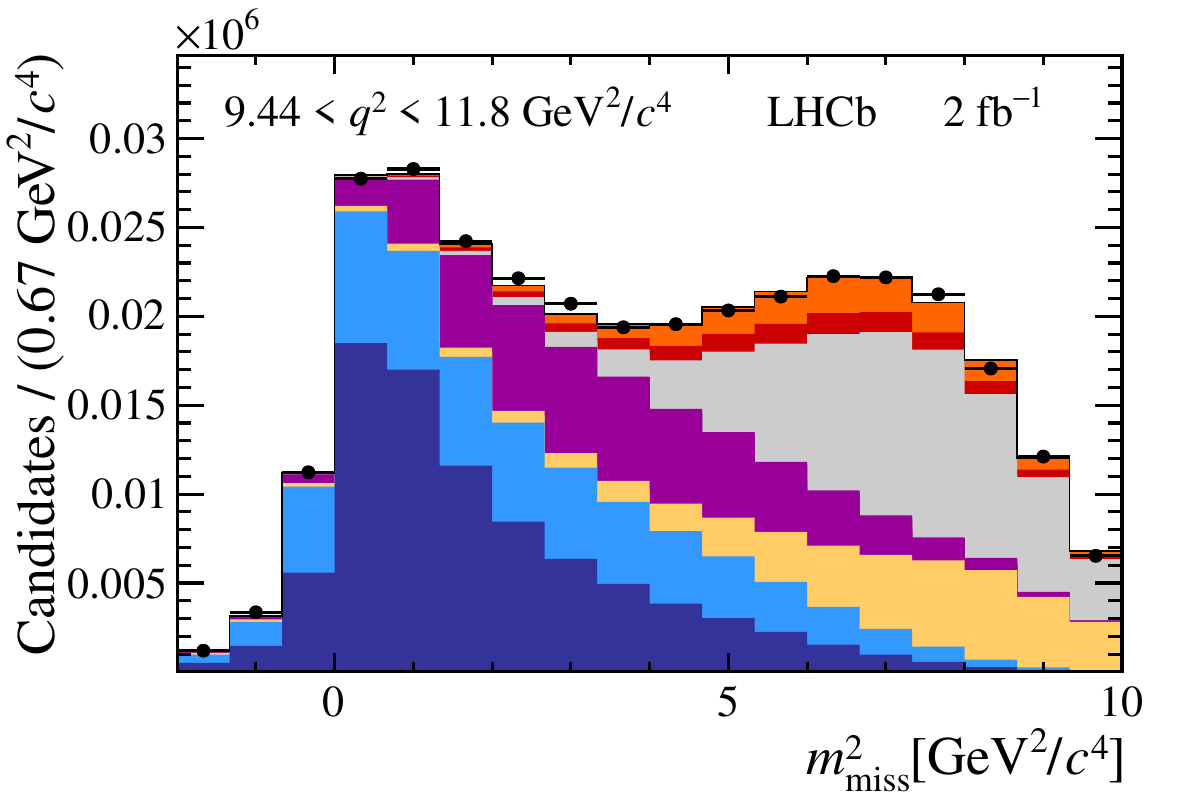}
    \includegraphics[width=0.45\textwidth]{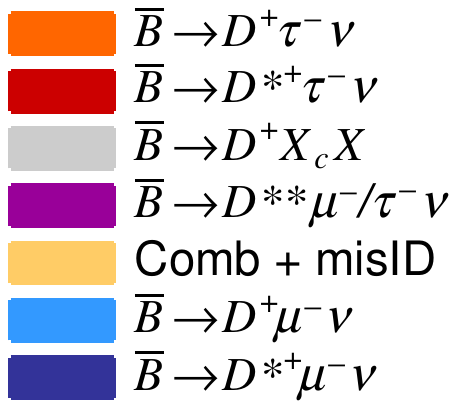}
    \includegraphics[width=0.45\textwidth]{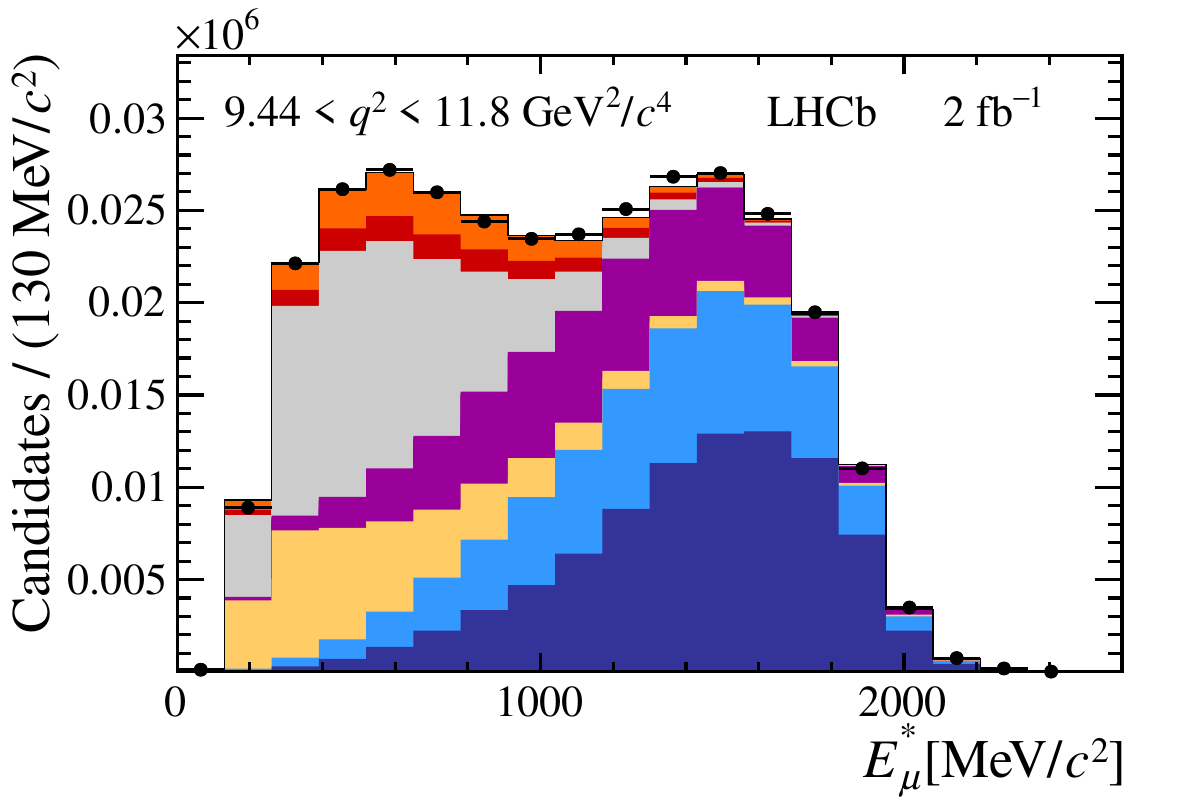}
    \caption{Distributions of the three kinematic variables in the signal isolation region, with the fit result overlaid. The $q^2$ distribution is shown over the full fit range whereas $m_{\rm miss}^2$ and $E_{\mu}^*$ are only shown in the range $9.44 < q^2 < 11.8 \aunit{Ge\kern -0.1em V}\xspace^2\!/c^4$.}
    \label{fig:Fit}
\end{figure}

One-dimensional projections of the three fit variables in the signal region are shown in Fig.~\ref{fig:Fit}, with the fit result overlaid. Signal yields of around 35000 and 29000 candidates are found for the decays \DpTauNu and \DstpTauNu, respectively.


The systematic uncertainties that affect this measurement are summarised in Table~\ref{tab:syst}. These uncertainties are dominated by those affecting the signal yield determination, and are therefore independent of the central values of $R(D^{(*)})$.

\begin{table}
\centering
\caption{Summary of systematic uncertainties on the $R(D^+)$ and $R(D^{*+})$ measurements. Systematic uncertainties associated with the efficiency are not shown as they are negligible.}
\label{tab:syst}
\begin{tabular}{l c c}
\hline			
Source      & $R(D^+)$ & $R(D^{\ast+})$ \\
\hline
Form factors & 0.023 & 0.035 \\
$\Bbar\to D^{**}[D^+ X] \mu/\tau \nu$ fractions & 0.024 & 0.025 \\
$\Bbar \to \Dp X_{c} X$ fraction & 0.020 & 0.034 \\
Misidentification & 0.019 & 0.012    \\
Simulation size & 0.009 & 0.030 \\
Combinatorial background & 0.005 & 0.020 \\
Data/simulation agreement & 0.016 & 0.011 \\
Muon identification & 0.008 & 0.027 \\
Multiple candidates & 0.007 & 0.017 \\
\hline
Total systematic uncertainty & 0.047 & 0.085\\
\hline
Statistical uncertainty       & 0.043 & 0.081 \\
\hline
\end{tabular}
\end{table}

The systematic uncertainty associated with the form factors that are allowed to vary in the fit is determined by fixing the form-factor parameters to their central values and assigning the quadratic difference in uncertainty of \RDp and \RDstp as a systematic.

The systematic uncertainty associated with the $\Bbar\to D^{**}\mun\neumb$ branching fractions of the individual $D^{**}$ states is obtained by fixing them to their best fit points and taking the quadratic difference in the uncertainties on \RDp and \RDstp to when they are constrained to the measurements quoted in Ref~\cite{Belle:2022yzd}. A similar method is also applied to the $B\to D^{**} \tau^- \bar{\nu}_{\tau}$ branching fraction constraint. The uncertainties on the $D^{**}\to \Dp X$ branching fractions are treated by introducing a set of nuisance parameters which control the abundance of charged pions in the different $D^{**}\to \Dp X$ decays. This ratio is varied in the fit and the difference is used to assign a systematic uncertainty.

In the baseline fit configuration, the fraction of the \Bub and \Bzb contributions to the $\Bbar\to \Dp X_{c} X$ background are fixed from simulation in both the signal and one-kaon isolation regions. This assumption is relaxed in an alternative fit by allowing them to vary in the fit and the difference in the results is assigned as a systematic uncertainty. In addition, a further categorisation is explored based on whether the $X_{c}$ meson is charged or neutral. The difference in the results when the background template is split into subsamples based on this categorisation is assigned as a systematic uncertainty.

Systematic uncertainties associated with the misidentification background arise from the treatment of fake tracks in the misidentification sample. Alternative definitions for fake tracks are explored and differences in the shapes are included as template shape variations in the fit. The treatment of the momentum smearing due to decays-in-flight of the hadron is also varied and included as an additional shape variation. Disabling these variations allows for a systematic to be determined based on the resulting uncertainties in the fit. An uncertainty on the assumption of the background in the PID calibration samples is determined by changing the procedure that accounts for the background in those samples.

The finite size of the simulated samples results in statistical uncertainties for each template. The effect of these on the results is determined by bootstrapping the templates and repeating the fit to the data. The variations of the central values are assigned as systematic uncertainties.

The combinatorial background shape is obtained from the same-sign sample with a multi-dimensional correction applied as a function of the visible mass and other kinematic variables. A systematic uncertainty is obtained by removing this correction and repeating the fit.

Potential differences between the data and simulation are investigated by removing the final simulation correction and repeating the fit.

The muon identification efficiency has a strong dependence on the muon momentum, which is different between the signal and normalisation modes. This efficiency is determined in bins of kinematic variables from a $\jpsi\to\mumu$ control sample. A systematic uncertainty is determined by increasing the number of bins by 20\% and repeating the measurement.

Approximately 2\% of the selected events contain multiple candidates, leading to a systematic uncertainty, which is determined by randomly selecting one candidate in those events and repeating the fit.

Finally, systematic uncertainties that were found to be negligible include the potential contribution from $\Bsb\to D^{**}_{s}\mun\neumb$ and $\Lb \to \Dp n \mun \neumb$ decays, the determination of the neutral isolation selection efficiency, the assumption that the reconstructed \Dp mass is uncorrelated with the fit variables in the \sPlot procedure and the effects of incomplete QED modelling in the simulation~\cite{deBoer:2018ipi}.


The ratios of the signal and normalisation yields are corrected for the relative efficiencies and the $\tau^{-} \to \mu^{-} \neumb \nu_{\tau}$ branching fraction~\cite{PDG2022}. This results in the following
\begin{align}\nonumber
    \RDp &= \RDpresult, \\
    \RDstp &= \RDstresult, \nonumber
\end{align}
where the first uncertainties are statistical and the second systematic. The correlation coefficient between the two measurements is \corr. These results are \compatibility from the average~\cite{HFLAV21} of SM predictions~\cite{Bigi:2016mdz,Gambino:2019sif,Bordone:2019vic,Bernlochner:2017jka,Jaiswal:2017rve,BaBar:2019vpl,Martinelli:2021onb}, $R(D)=0.298\pm0.004$ and $R(D^*)=0.254\pm0.005$, and 1.09$\,\sigma$ from the world average~\cite{HFLAV21}. These are the first measurements of \RDp and \RDstp using the $\Dp\to \Km\pip\pip$ decay mode at \lhcb, the first analysis which uses tracker-only simulation and the first measurement to use HAMMER during the minimisation procedure of the likelihood fit. Assuming isospin symmetry between the charged and neutral decays, a combination with other \lhcb measurements~\cite{LHCb-PAPER-2022-052,LHCb-PAPER-2022-039} results in $R(D)=0.335 \pm 0.052$ and $R(D^{\ast})= 0.279 \pm 0.019$, with a correlation coefficient of $-0.30$. This combination is also consistent with the SM prediction. Contributions to the systematic uncertainties due to sources common across the measurements are conservatively assumed to be 100\% correlated. These contributions include the uncertainty associated with the form factors and branching fractions constrained from external measurements. The precision of these measurements is primarily limited by the size of the signal and control samples, so more precise measurements are expected with future \lhcb datasets.

\section*{Acknowledgements}
%
%
\noindent We would like to thank Dean Robinson and Michele Papucci for extensive discussions and technical advice on the use of the HAMMER tool. We express our gratitude to our colleagues in the CERN
accelerator departments for the excellent performance of the LHC. We
thank the technical and administrative staff at the LHCb
institutes.
We acknowledge support from CERN and from the national agencies:
CAPES, CNPq, FAPERJ and FINEP (Brazil); 
MOST and NSFC (China); 
CNRS/IN2P3 (France); 
BMBF, DFG and MPG (Germany); 
INFN (Italy); 
NWO (Netherlands); 
MNiSW and NCN (Poland); 
MCID/IFA (Romania); 
MICINN (Spain); 
SNSF and SER (Switzerland); 
NASU (Ukraine); 
STFC (United Kingdom); 
DOE NP and NSF (USA).
We acknowledge the computing resources that are provided by CERN, IN2P3
(France), KIT and DESY (Germany), INFN (Italy), SURF (Netherlands),
PIC (Spain), GridPP (United Kingdom), 
CSCS (Switzerland), IFIN-HH (Romania), CBPF (Brazil),
and Polish WLCG (Poland).
We are indebted to the communities behind the multiple open-source
software packages on which we depend.
Individual groups or members have received support from
ARC and ARDC (Australia);
Key Research Program of Frontier Sciences of CAS, CAS PIFI, CAS CCEPP, 
Fundamental Research Funds for the Central Universities, 
and Sci. \& Tech. Program of Guangzhou (China);
Minciencias (Colombia);
EPLANET, Marie Sk\l{}odowska-Curie Actions, ERC and NextGenerationEU (European Union);
A*MIDEX, ANR, IPhU and Labex P2IO, and R\'{e}gion Auvergne-Rh\^{o}ne-Alpes (France);
AvH Foundation (Germany);
ICSC (Italy); 
GVA, XuntaGal, GENCAT, Inditex, InTalent and Prog.~Atracci\'on Talento, CM (Spain);
SRC (Sweden);
the Leverhulme Trust, the Royal Society
 and UKRI (United Kingdom).

\addcontentsline{toc}{section}{References}
\bibliographystyle{LHCb}
\bibliography{main,standard,LHCb-PAPER,LHCb-CONF,LHCb-DP,LHCb-TDR}

\newpage

\centerline
{\large\bf LHCb collaboration}
\begin
{flushleft}
\small
R.~Aaij$^{36}$\lhcborcid{0000-0003-0533-1952},
A.S.W.~Abdelmotteleb$^{55}$\lhcborcid{0000-0001-7905-0542},
C.~Abellan~Beteta$^{49}$,
F.~Abudin{\'e}n$^{55}$\lhcborcid{0000-0002-6737-3528},
T.~Ackernley$^{59}$\lhcborcid{0000-0002-5951-3498},
A. A. ~Adefisoye$^{67}$\lhcborcid{0000-0003-2448-1550},
B.~Adeva$^{45}$\lhcborcid{0000-0001-9756-3712},
M.~Adinolfi$^{53}$\lhcborcid{0000-0002-1326-1264},
P.~Adlarson$^{79}$\lhcborcid{0000-0001-6280-3851},
C.~Agapopoulou$^{13}$\lhcborcid{0000-0002-2368-0147}
% ...

{\footnotesize \it

% ...
$^{13}$Universit{\'e} Paris-Saclay, CNRS/IN2P3, IJCLab, Orsay, France\\
% ...
$^{36}$Nikhef National Institute for Subatomic Physics, Amsterdam, Netherlands\\
% ...
$^{45}$Instituto Galego de F{\'\i}sica de Altas Enerx{\'\i}as (IGFAE), Universidade de Santiago de Compostela, Santiago de Compostela, Spain\\
% ...
$^{49}$Physik-Institut, Universit{\"a}t Z{\"u}rich, Z{\"u}rich, Switzerland\\
% ...
$^{53}$H.H. Wills Physics Laboratory, University of Bristol, Bristol, United Kingdom\\
% ...
$^{55}$Department of Physics, University of Warwick, Coventry, United Kingdom\\
% ...
$^{58}$School of Physics and Astronomy, University of Glasgow, Glasgow, United Kingdom\\
$^{59}$Oliver Lodge Laboratory, University of Liverpool, Liverpool, United Kingdom\\
% ...
$^{67}$Syracuse University, Syracuse, NY, United States\\
% ...
$^{79}$Department of Physics and Astronomy, Uppsala University, Uppsala, Sweden, associated to $^{58}$\\
% ...
}
\end{flushleft}

\end{document}